\documentclass[journal]{IEEEtran}
\usepackage{amsmath,amsfonts}
\usepackage{algorithmic}
\usepackage{algorithm}
\usepackage{array}
\usepackage[caption=false,font=normalsize,labelfont=sf,textfont=sf]{subfig}
\usepackage{textcomp}
\usepackage{stfloats}
\usepackage{url}
\usepackage{verbatim}
\usepackage{graphicx}
\usepackage{cite}
\begin{document}

\title{50 mm $\times$ 50 mm Cesium Atomic Vapor Cell for Terahertz Imaging: Implementation and Application }

\author{Bin Zhang, Jun Wan, Tao Li, Xian-Zhe Li, Yu Wu, Qi-Rong Huang, Xin-Yu Yang, Wei Huang, 
Kai-Qing Zhang, and Hai-Xiao Deng~\IEEEmembership{}
\thanks{This work was supported by the CAS Project for Young Scientists in Basic Research (YSBR-042), the National Natural Science Foundation of China (12125508, 12105347, and 12275340), Innovation Program of Shanghai Advanced Research Institute, CAS (2024CP001), and Basic Research Program of Science and Technology Commission of Shanghai Municipality(25ZR1402527). $\textit{(Corresponding author:Haixiao Deng, Kaiqing Zhang.)}$

Bin Zhang is with the Shanghai Advanced Research Institute, Chinese Academy of Sciences and also with the ShanghaiTech University, Shanghai 201210, China.
 
Jun Wan, Tao Li, Xian-Zhe Li, Yu Wu, Kai-Qing Zhang, and Hai-Xiao Deng are with the Shanghai Advanced Research Institute, Chinese Academy of Sciences, Shanghai 201210, China(e-mail: zhangkq@sari.ac.cn, denghx@sari.ac.cn).

Qi-Rong Huang, Xin-Yu Yang, and Wei Huang are with the Laboratory of Quantum Engineering and Quantum Materials, School of physics, South China Normal University, 
Guangzhou 510631, China.}}



\maketitle

\begin{abstract}
Rydberg atomic sensors offer transformative potential for high-speed, high-sensitivity terahertz (THz) imaging. However, previous systems are hindered by restricted imaging areas, largely due to the compact dimension of atomic vapor cells and inefficient beam-shaping methodologies. We present a THz imaging system with a 50 mm $\times$ 50 mm area, enabled by a custom-engineered scaled-up atomic vapor cell and an optimized beam-shaping optical architecture. Experimental validation confirms that this system achieves near-diffraction-limited, high-resolution THz imaging at 0.55 THz under ambient conditions. Furthermore, its capabilities are demonstrated through real-time visualization of the diffusion dynamics of a deionized water droplet in anhydrous ethanol. This work not only expands the boundaries of Rydberg atomic sensors but also establishes a critical foundation for advancing THz imaging technologies toward into real-world, large-scale applications.
\end{abstract}


\section{Introduction}
\IEEEPARstart{T}{erahertz} (THz) radiation, spanning frequencies from 0.1 to 10 THz, occupies the spectral region between microwaves and infrared light. Thanks to its unique ability to penetrate non-conductive materials such as plastics and ceramics, while maintaining photon energies several orders of magnitude lower than those of X-rays, THz radiation has become a powerful tool for non-destructive imaging 
\cite{leitenstorfer2023TerahertzScience2023}
\cite{jansenTerahertzImagingApplications2010a}
\cite{mittlemanTwentyYearsTerahertz2018}.
In contrast to conventional optical imaging, which primarily captures structural information, THz radiation offers both sub-millimetre spatial resolution and spectroscopic fingerprints of molecular vibrations, such as hydrogen bonding in biomolecules\cite{falconerTerahertzSpectroscopicAnalysis2012}\cite{jepsenTerahertzSpectroscopyImaging2011}. 
These distinctive features have enabled transformative applications in areas such as security screening\cite{federiciReviewTerahertzSubterahertz2010}\cite{liuTerahertzSpectroscopyImaging2007}, biomedical diagnostics
\cite{yangBiomedicalApplicationsTerahertz2016}
\cite{brunTerahertzImagingApplied2010}
\cite{zaytsevHighlyAccurateVivo2015}, and industrial quality control
\cite{ahiAdvancedTerahertzTechniques2016}
\cite{gowenTerahertzTimeDomain2012}
\cite{wietzkeTerahertzImagingNew2007}. The growing demand for high-resolution, high-sensitivity, and non-ionizing imaging technologies highlights the strategic importance of advancing THz systems.

The development of highly sensitive THz detectors is an important goal in THz science and technology. Current THz band detectors can be broadly classified as thermal, electronic or photonic\cite{koppensPhotodetectorsBasedGraphene2014}
\cite{vicarelliGrapheneFieldeffectTransistors2012}
\cite{lewisReviewTerahertzDetectors2019}. Thermal detectors primarily measure the energy released when THz photons are absorbed. The Golay cell, for example, can operate at room temperature, has a simple structure, but limited sensitivity\cite{rogalskiTerahertzDetectorsFocal2011}\cite{yamashitaMiniaturizedInfraredSensor1998}. Bolometers can achieve high sensitivity and low-noise detection, but they require operation at low temperatures\cite{hargreavesTerahertzImagingMaterials2007}. Microbolometer arrays operating at room temperature can also perform terahertz measurements at 25 frames per second (FPS) \cite{odenImagingBroadbandTerahertz2013}. Pyroelectric detectors have a relatively simple structure and can operate at room temperature, but they have a slow response time and high internal thermal noise. They are also less sensitive and deteriorate at longer wavelengths \cite{kuznetsovSelectivePyroelectricDetection2016}. Electronic detectors typically use measurements of electron-THz photon interactions to detect THz signals. Examples include Schottky diodes, graphene field-effect transistors, resonant tunneling diodes and other semiconductor diodes \cite{mehdiTHzDiodeTechnology2017,
yadavStateArtRoomTemperature2023,
feiginovFrequencyLimitationsResonantTunnelling2019,
vitielloRoomTemperatureTerahertzDetectors2012}. These detectors can operate at room temperature, can achieve higher sensitivity and faster response times than thermal detectors. Howerver, they have limited bandwidth and low sensitivity at high frequencies \cite{otsujiTrendsResearchModern2015}. Photonic detectors are sensitive to optical signals in the THz field. They achieve THz field probing by mixing THz signals with different reference signals. However, they are usually narrow-banded \cite{gregoryContinuouswaveTerahertzSystem2005}
\cite{siebertContinuouswaveAlloptoelectronicTerahertz2002}
\cite{taimreLaserFeedbackInterferometry2015}. While these detectors offer various possibilities for THz detection imaging, they have yet to meet the current demands of the field for high-speed, high-sensitivity detection at room temperature. In recent years, THz detection techniques based on Rydberg atoms have introduced new concepts to the field. Rydberg atoms are a class of atoms in which electrons are excited to atomic states with high principal quantum numbers. Compared with atoms in the ground state or low-excited states, Rydberg atoms possess huge electric dipole transitions and are highly sensitive to weak THz electric fields, enabling high-sensitivity detection \cite{schleichQuantumTechnologyResearch2016,adamsRydbergAtomQuantum2020,chenTerahertzElectrometryInfrared2022
,sheRydbergAtomTerahertzHeterodyne2024a,jingAtomicSuperheterodyneReceiver2020}.

Previous studies have demonstrated the feasibility of THz-to-optical conversion using Rydberg atoms \cite{wadeRealtimeFieldTerahertz2017}. More recently, a full-field, real-time THz imaging system based on Rydberg atoms was reported \cite{downesFullFieldTerahertzImaging2020}. This system achieves THz-to-visible frequency conversion through the interaction between Rydberg-state atomic vapor and THz radiation, enabling image information to be captured via a visible-light camera. Theoretically, such systems can support video-rate imaging at the MHz level. Additionally, a dual-mode THz imaging system capable of both highly sensitive THz detection and imaging frame rates up to 6000 FPS has been developed \cite{liDualcamerasTerahertzImaging2025}. 
Although THz imaging systems based on Rydberg atoms show considerable promise, several technical challenges remain. For instance, high noise levels and limited imaging quality have been observed. To mitigate these issues, deep learning-based approaches have been proposed\cite{wanEnhancingTerahertzImaging2025}, which significantly enhance image quality and signal robustness. However, other limitations persist, including the complexity of fabricating large-area atomic vapor cells, the difficulty of precise temperature control, and inefficient laser energy utilization. Consequently, such systems often suffer from a limited imaging area and non-uniform fluorescence intensity distribution. These constraints currently hinder the broader practical deployment of Rydberg atom-based THz imaging technologies.

To overcome the aforementioned limitations, we extended the length and width of the cesium (Cs) atomic vapor cell while minimizing its thickness, thereby increasing the imaging area and reducing internal THz field reflections. A combination of Powell prisms and cylindrical lenses was introduced to shape the laser beam uniformly across the expanded cell area. THz imaging experiments confirmed that the atomic sensor achieved a 50 mm $\times$ 50 mm effective detection area at room temperature, with a spatial resolution of 1.25 mm under a 0.55 THz field—approaching the system’s diffraction limit. Furthermore, we investigated the dripping and diffusion dynamics of two colorless liquids, water and ethanol. As ethanol–water mixtures are simple amphiphilic systems often used as models for studying molecular interactions in biological processes\cite{zhaoMolecularDynamicInvestigation2021}\cite{chakrabortyHydrogenBondStructure2023}, we recorded the temporal evolution of the free diffusion interface between the two. These results suggest that the proposed approach may provide a valuable tool for visualizing and analyzing dynamic processes in chemical or biological systems with distinct THz absorption characteristics.

\begin{figure}[H] 
\centering
\includegraphics[width=1.0\linewidth, height=0.2\textheight]{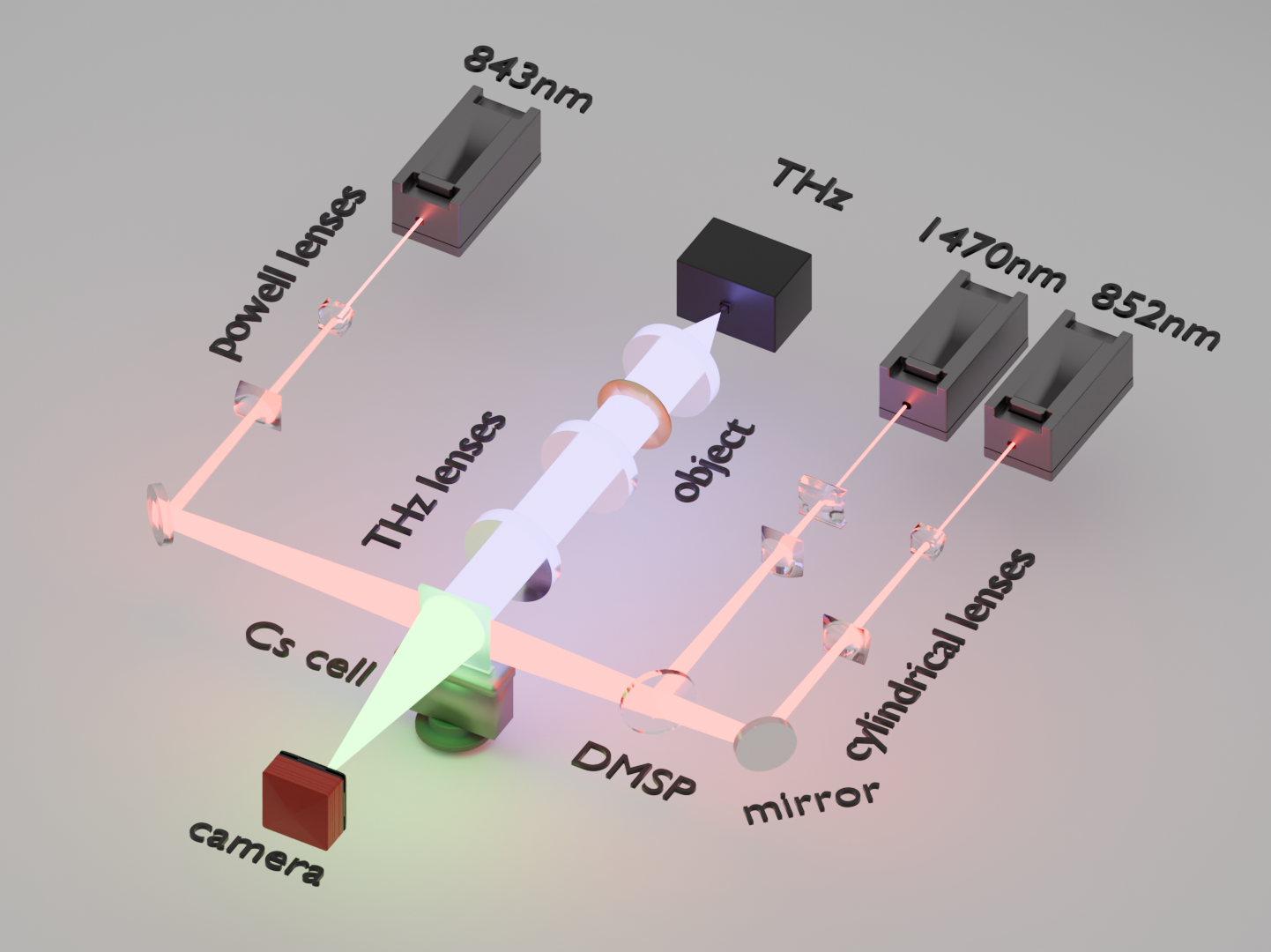} 
\label{1}
\caption{Experiment layout. The experimental setup includes a 0.55 THz source, narrow-linewidth semiconductor lasers with central wavelengths of 843 nm, 1470 nm, and 852 nm respectively, a Cs atomic vapor cell, and a camera. DMSP is a shortpass dichroic mirror.} 
\end{figure}
\section{Experimental setup}
\subsection{System Overview}
Fig. 1 illustrates the schematic of the experimental setup. Three infrared laser beams are frequency-stabilized and spatially shaped before being directed into the atomic vapor cell, where Cs atoms are excited from the ground state to a Rydberg state. After traversing the imaging optical path, the incident THz radiation interacts with the Rydberg atoms, enabling frequency conversion from THz to visible light. In the experiment, the probe laser (852 nm,45 mW) is stabilized to the $6S_{1/2}F=4 \rightarrow 6P_{3/2}F^{\prime}= 5$ hyperfine transition using saturated absorption spectroscopy. The coupling laser (1470 nm, 25 mW) is frequency-stabilized to the $6P_{3/2}F^{\prime}=5\rightarrow7S_{1/2}F^{\prime}=4$ transition via electromagnetically induced transparency. The Rydberg laser (843 nm, 450 mW) is tuned to the 7$S_{1/2}\rightarrow 14P_{3/2}$ transition to complete the excitation to the Rydberg state. When irradiated by a resonant THz field at 548.613 GHz with a power of 1.85 mW, the atoms will shift from the $14P_{3/2}$ state to the $13D_{5/2}$ state and emit 535 nm photons of visible light outward\cite{downesFullFieldTerahertzImaging2020}.

\subsection{Atomic Vapor Cell Design}
The atomic vapor cell used in the system is a rectangular hollow structure with external dimensions of 60 mm $\times$ 100 mm $\times$ 5 mm and a wall thickness of 2 mm. For practical imaging, the effective internal vapor region is approximately 50 mm $\times$ 50 mm $\times$ 1 mm. The cell is made of optically transparent quartz glass to facilitate beam transmission and collection, and it is evacuated to a vacuum level of 5 × 10$^{-4}$ Pa. Approximately 10 mg of Cs was sealed inside the cell. The vapor cell is housed within a 40 mm deep Teflon enclosure and securely mounted onto a metal base for mechanical stability and optical alignment. The Teflon enclosure was originally designed to conduct heat and facilitate the heating of the atomic vapor cell. However, in the current experiments, all measurements were performed at room temperature, and no external heating was applied to the vapor cell.
\begin{figure*}[htp]
\begin{center}
\includegraphics[width=0.95\textwidth]{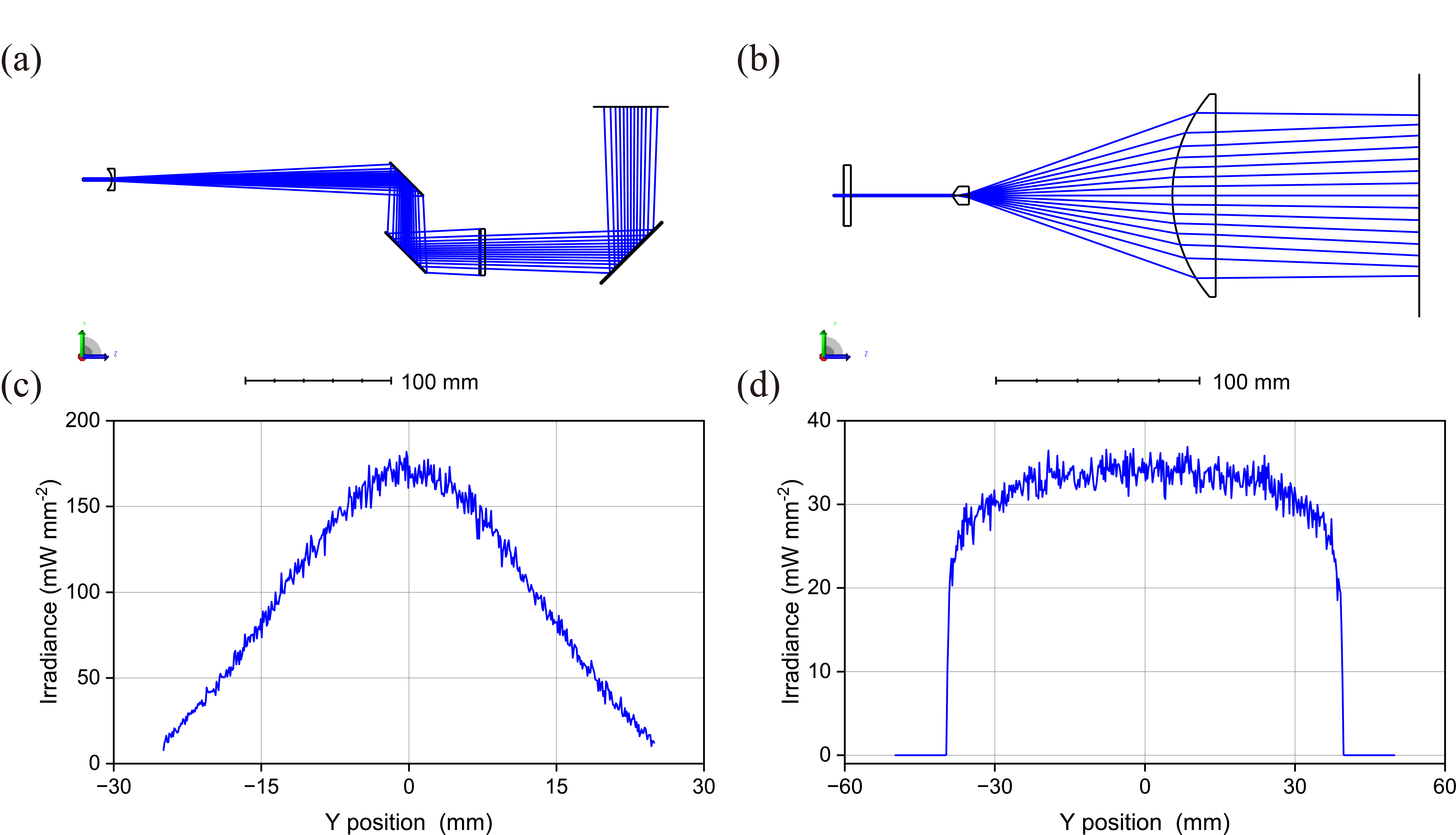}
\end{center}
\caption{(a) Simulated 3D layout of the cylindrical lens set. The mirror is used to collapse the optical path. (b) Simulated 3D layout of the Powell prism. The beam diameter is compressed before incidence to the optimal use range of the Powell prism, and the cylindrical plano-convex lenses are used for unidirectional beam compression as well as beam collimation. (c) Y-direction cross-section illuminance distribution of the cylindrical lens set. (d) Y-direction cross-section illuminance distribution of the Powell prism. The illuminance distributions are taken at the centre of the X-axis of the image plane.\label{2}}
\end{figure*}
\subsection{Simulation and Implementation of Beam Shaping}
To generate a fluorescence sheet that matches the dimensions of the large vapor cell, the laser beam must be shaped into a linearly extended spot of appropriate size. In prior work, this was achieved by converting the circular infrared beam into a linear profile using a pair of orthogonally oriented cylindrical lenses. These lenses focus or expand the beam in a single axis, enabling the formation of a narrow, elongated spot at the location of the vapor cell through the overlap of the focused axes. Under paraxial approximation, the relationship between the beam expansion width \textit{L} and the focal length \textit{f} of the cylindrical lens is given by:
\begin{equation}
L = 2(\frac{r}{f})(z+f)
\end{equation}
where $r$ is the radius of the incident beam (neglecting divergence in the Gaussian profile), and $z$ is the image distance. To obtain the desired length and width of the laser line spot at position $z$, an appropriate combination of plano-concave and plano-convex cylindrical lenses should be selected. However, the laser beam used in the experiment exhibits a non-uniform intensity distribution. Although cylindrical lenses can expand or compress the beam in a single axis, they do not modify the inherent power distribution. As a result, the generated two-dimensional fluorescence sheet shows spatial inhomogeneity, which limits the uniformity of the imaging across the vapor cell.
\begin{figure}[htp] 
\centering
\includegraphics[width=1\linewidth]{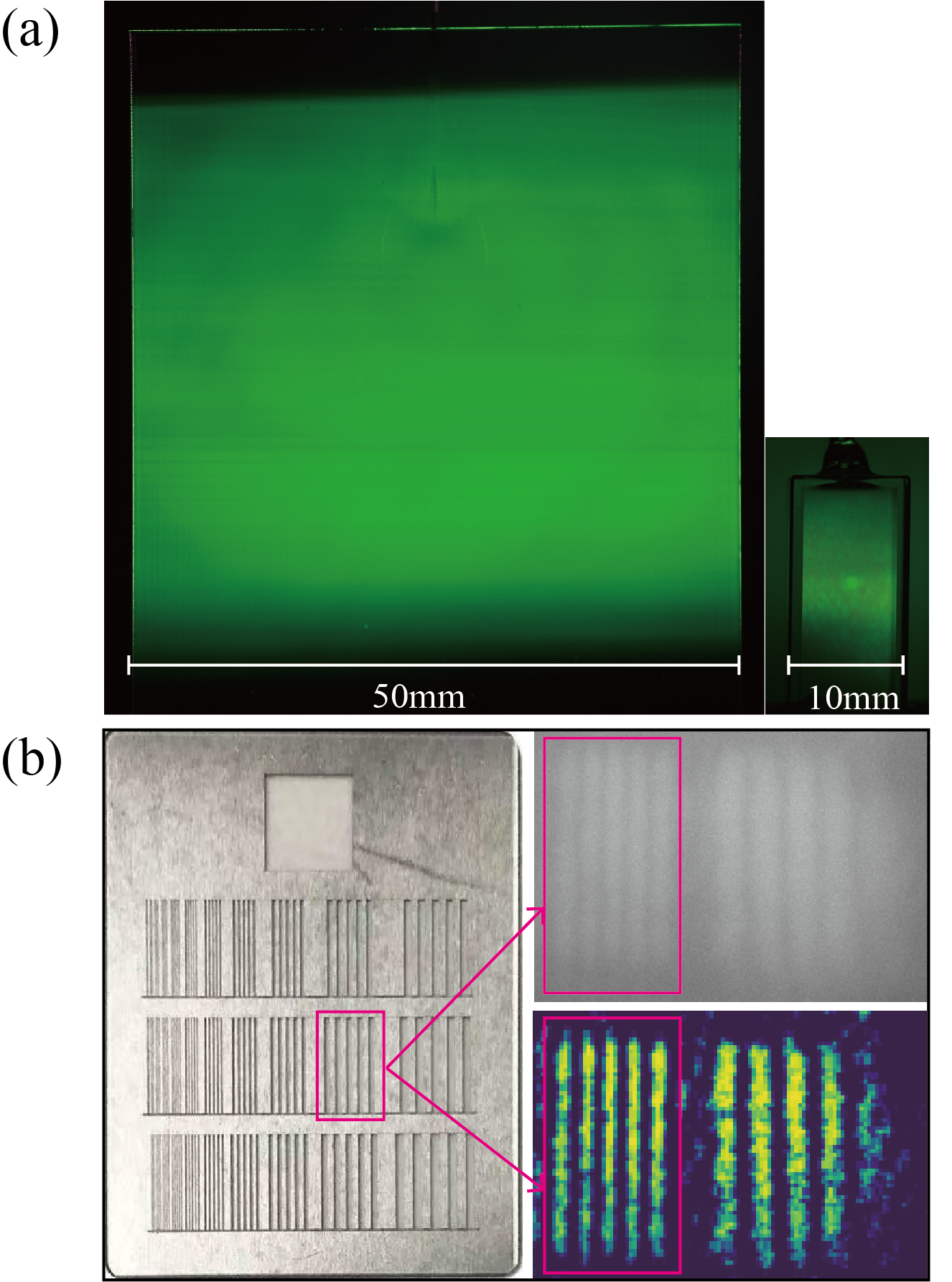}
\label{3}
\caption{(a) Comparison of THz atomic fluorescence images obtained using the large-area atomic vapor cell in this study and a 20 mm $\times$ 10 mm cell. The image corresponding to the proposed system was captured with a Sony A7M4 camera under the following settings: exposure time 1/2 s, aperture f/2.8, and ISO 2000. (b) Spatial resolution test using a German IBA TYPE18D line pair card. The physical image of the test card is shown on the left. The original THz fluorescence image (top right) was acquired using an sCMOS camera, while the corresponding false-color image (bottom right) was generated through contrast-enhancement algorithms. The sCMOS camera was operated in high-sensitivity gain mode with an exposure time of 200 ms.} 
\end{figure}

Assuming no power loss of the beam as it passes through the lens, there is following equation:
\begin{equation}
\int2{\pi}I_{in}(\frac{r_{1}}{\omega_{0}})r_{1}dr_{1}=\int2{\pi}I_{out}(\frac{r_{2}}{z_{0}})r_{2}dr_{2}
\end{equation}
Here, $r_{1}$ denotes the projection height of the incident surface, and $\omega_{0}$ is the waist radius of the incident beam. $r_{2}$ represents the projection height at the output surface, while $z_{0}$ is the beam radius after propagation. $I_{in}$ and $I_{out}$ are the intensities of the incident and outgoing laser beams, respectively. By appropriately selecting the parameters of the incident beam, the projection height and intensity distribution at the output surface can be controlled to achieve the desired output beam profile. Commercial Powell prisms provide a well-established aspherical solution for transforming a Gaussian beam of a specific wavelength and spot size into a linear beam with uniform intensity distribution. To evaluate the beam-shaping performance of cylindrical lenses and Powell prisms, optical simulations were conducted using Ansys Zemax OpticStudio. The simulation results are presented in Fig. 2. In Fig. 2, (a) and (b) display the 3D optical layouts of the cylindrical lens pair and the Powell prism–cylindrical lens combination, respectively, (c) and (d) show the normalized Y-direction cross-section illuminance distribution for each configuration. The simulation parameters are defined as follows: For Fig. 2(a) , the system entrance pupil diameter is 2 mm, the plano-concave cylindrical lens focal length is –20 mm, the plano-convex cylindrical lens focal length is 100 mm. For Fig. 2(b) , the system entrance pupil diameter is 0.8 mm, the divergence angle of Powell prism is 30$^\circ$, the plano-convex cylindrical lenses focal lengths are 300 mm and 150 mm, respectively. The common parameters of both are: the apodization type is Gaussion, the apodization factor is 1, the wavelength is 852 nm.
\begin{figure}[htp] 
\centering
\includegraphics[width=1\linewidth]{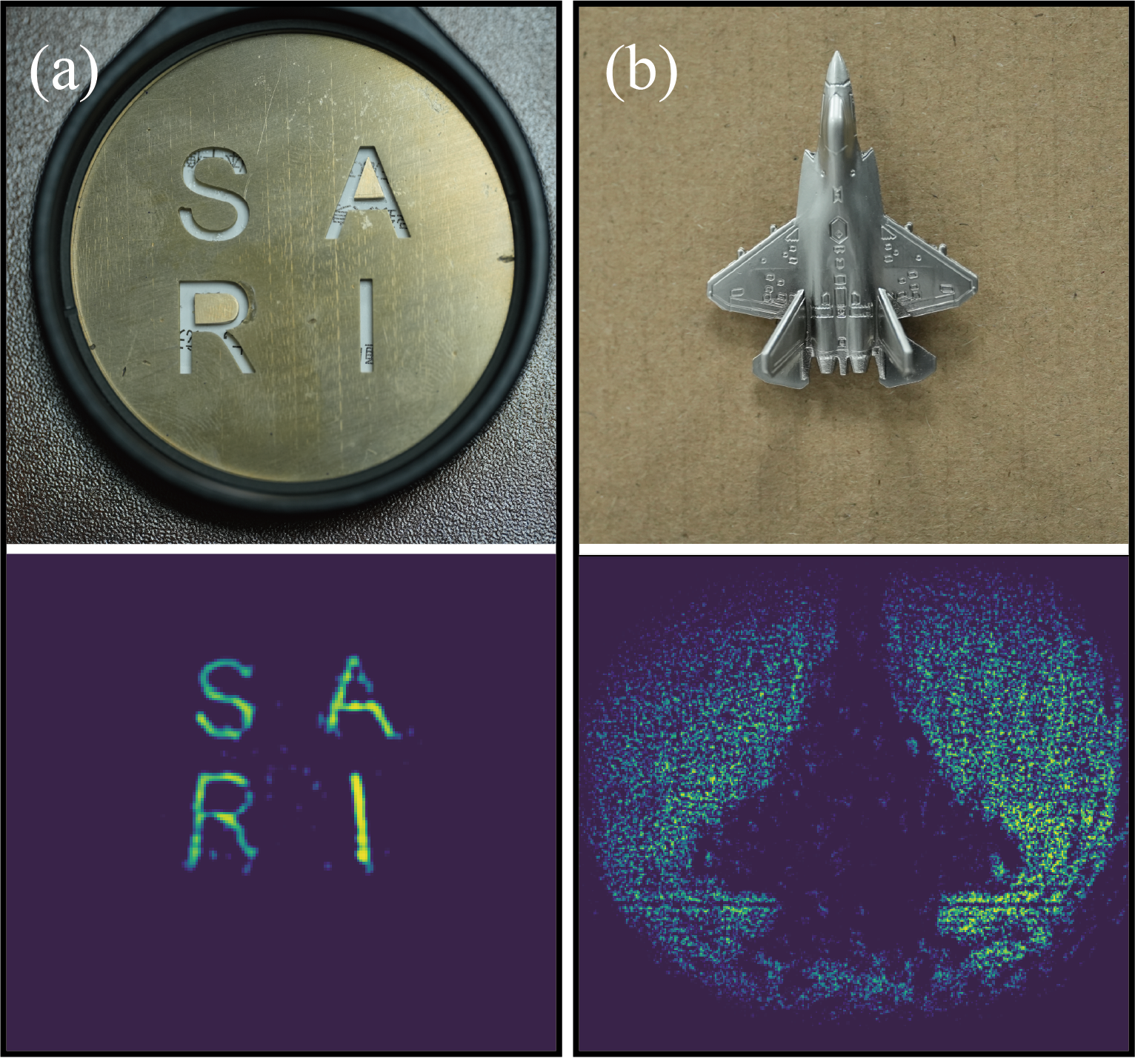}
\label{4}
\caption{(a) Photograph of the brass plate engraved with the letters ``SARI'' (top) and the corresponding THz atomic fluorescence image after algorithmic contrast enhancement (bottom). (b) Photograph of a zinc-alloy aircraft model mounted on cardboard (top) and its corresponding algorithmically enhanced THz fluorescence image (bottom).} 
\end{figure}

As shown in Fig. 2, the laser energy distribution becomes significantly more uniform after beam shaping with a Powell prism. Therefore, in the experiment, a combination of a Powell prism (fan angle 30$^\circ$) and a plano-convex cylindrical lens (focal length = 50 mm) was employed to shape and homogenize the 852 nm and 843 nm infrared laser beams into linear spots. Since Powell prisms are highly sensitive to the diameter of the incident beam, optimal shaping requires a beam diameter of approximately 0.8 mm. To achieve this, a Keplerian beam reduction system was implemented to compress the laser spot before it enters the Powell prism. However, the shaped beam exhibits a large divergence angle and expands rapidly along the line direction, necessitating the use of a plano-convex cylindrical lens for collimation. For the 1470 nm laser, due to the unavailability of compatible Powell prisms in this wavelength range, a conventional shaping scheme was adopted using a combination of cylindrical lenses. After initial collimation, the beam passes sequentially through a plano-concave cylindrical lens (focal length = -30 mm), a Dove prism, mirrors, and a plano-convex cylindrical lens (focal length = 400 mm) before reaching the atomic vapor cell. The two cylindrical lenses are oriented orthogonally to expand the beam vertically and compress it horizontally. Fine adjustments to the Dove prism and mirrors are used to control the position and incident angle of the resulting linear spot on the vapor cell.

\subsection{THz Fluorescence Detection}
The three laser beams are shaped into linear spots, directed by mirrors, and made to converge at the atomic vapor cell to form a thin fluorescence sheet of excited atoms. The THz source is positioned behind the cell, emitting radiation that propagates along the vertical axis of the vapor cell. This radiation is collimated using a Teflon lens (Thorlabs LAT150) and directed onto the target object. The transmitted THz wave then passes through a THz imaging lens (Huirui Optics F0.7s), projecting the object’s spatial information onto the fluorescence sheet within the cell. The resulting fluorescence, generated through the interaction between the THz radiation and Rydberg-excited atoms, is recorded by imaging cameras. True-color fluorescence images were captured using a Sony A7M4 digital camera equipped with a Sigma 105 mm F2.8 DG DN MACRO lens. For high-resolution and low-noise imaging, a Dhyana 400 BSI V3 sCMOS camera paired with a cemented doublet lens (focal length = 50 mm) was used. To enhance the signal-to-noise ratio, a narrow band-pass optical filter (central wavelength = 532 nm, bandwidth = 10 nm, OD 4) was placed in front of the imaging camera.

\section{Results and discussion}
\subsection{THz Fluorescence Imaging and Analysis}
Fig. 3(a) presents a comparison between THz-induced atomic fluorescence images obtained using a 20 mm $\times$ 10 mm vapor cell and those captured with the larger vapor cell implemented in this system. The fluorescence region generated in our system spans approximately 50 mm in both length and width, with a thickness of around 100 \textmu m. These results confirm that the system enables a significantly larger imageable area and exhibits improved fluorescence uniformity. Although the optimized laser beam-shaping scheme effectively mitigates the non-uniformity in laser power distribution, it does not eliminate it completely. The laser power density directly influences the fluorescence intensity of the excited atoms, which in turn impacts the signal-to-noise ratio (SNR) of the imaging. To ensure sufficient SNR, the atomic fluorescence must be strong enough; otherwise, the signal may be overwhelmed by noise and rendered undetectable. 
\begin{figure*}[htp]
\begin{center}
\includegraphics[width=0.95\textwidth]{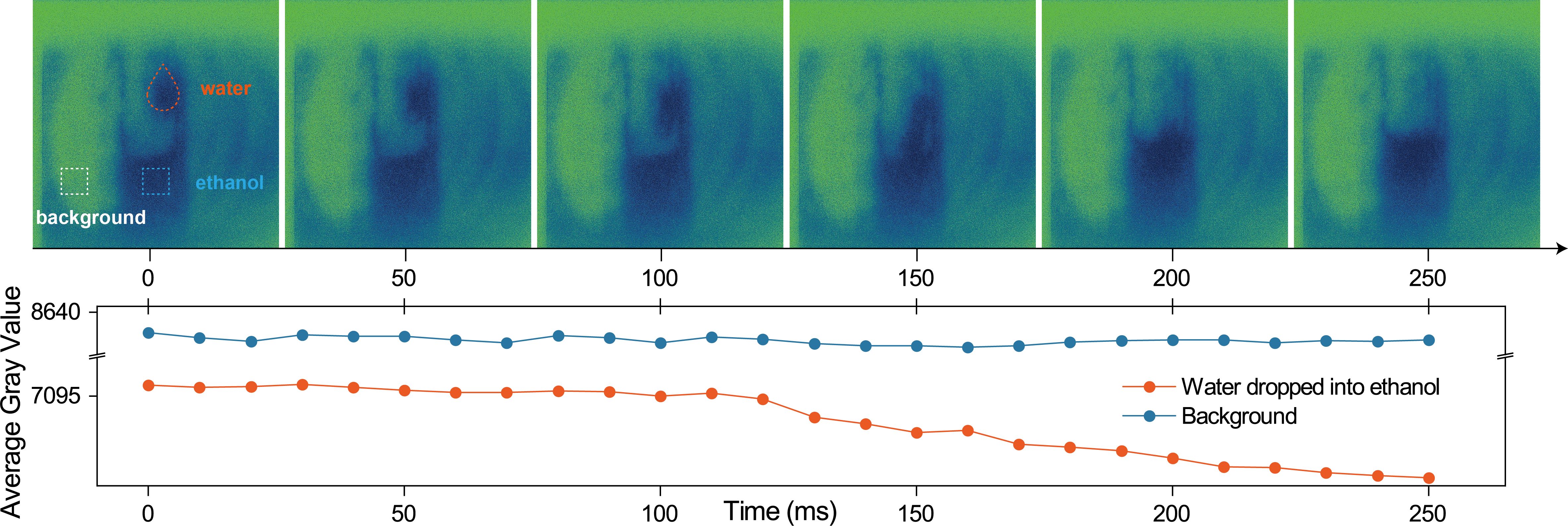}
\end{center}
\caption{Liquid mixing experiment. Visualization of the mixing process as deionized water drips into anhydrous ethanol, captured at 100 FPS with an exposure time of 10 ms per frame. For clarity, a frame is extracted every 50 ms in this figure. In the images, the red dashed line indicates the region of deionized water, the blue dashed line indicates the region of anhydrous ethanol, and the white dashed line denotes the background reference region. The lower panel presents the temporal evolution of the average gray values in the ethanol and background regions, calculated from the blue-boxed (ethanol) and white-boxed (background) areas, respectively. Full-frame dynamic image sequences and corresponding GIFs are provided in the supplementary material.\label{5}}
\end{figure*}

A previous study demonstrated that spatial resolution near the diffraction limit could be achieved at approximately 1 mm\cite{downesFullFieldTerahertzImaging2020}. To evaluate the spatial resolution of our system, we performed an imaging test using a Type 18D wire-pair resolution card. Fig. 3(b) displays a photograph of the test card, the raw image captured by the sCMOS camera, and the post-processed imaging result. At the location indicated by the arrows, the wire-pair density is 0.8 line pairs per millimetre (lp/mm), and each vertical line can be clearly resolved. This corresponds to a spatial resolution of 1.25 mm. In this experiment, the wavelength of the incident THz radiation is approximately 0.55 mm. The imaging lens has an aperture \textit{D} = 42.5 mm, and the object distance is \textit{L} = 60 mm. According to the Rayleigh criterion, the angular resolution is given by:
\begin{equation}
\theta = 1.22(\frac{\lambda}{D})
\end{equation}
The corresponding spatial resolution at distance \textit{L} is:
\begin{equation}
\Delta x = L \times \theta
\end{equation}
We can calculate $\Delta x$ = 0.947 mm. Although this imaging system is not an ideal diffraction-limited optical system, the experimentally measured resolution of 1.25 mm closely approaches the theoretical limit of 0.947 mm. This indicates that the system has achieved near-diffraction-limited performance under current experimental conditions.
\begin{figure}[htp] 
\centering
\includegraphics[width=1\linewidth]{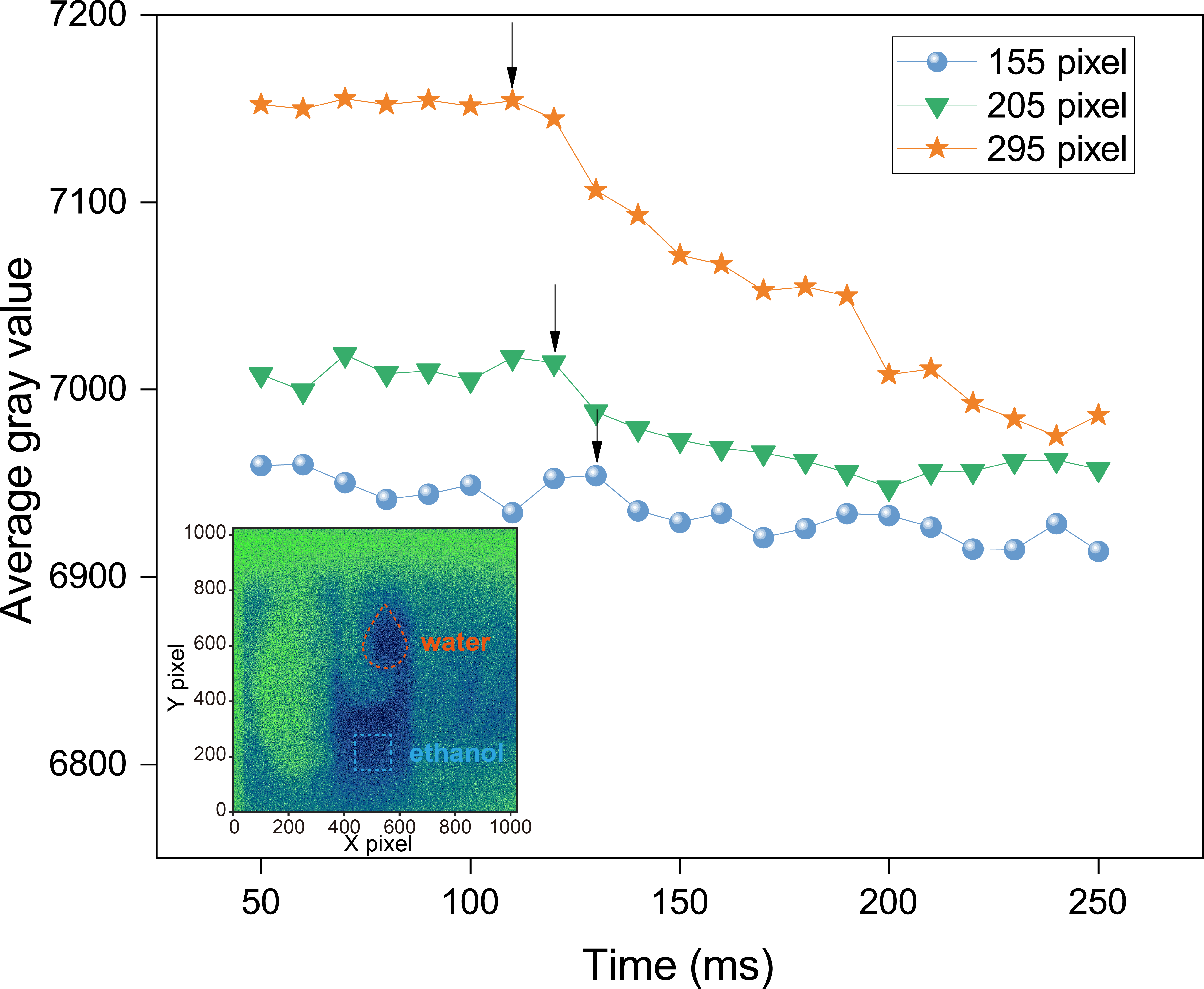}
\label{6}
\caption{Temporal evolution of average gray values at different vertical positions beneath the liquid surface. Each curve corresponds to a specific depth within the ethanol solution. Black arrows indicate the moments when water reaches the corresponding positions, resulting in a noticeable decrease in gray value due to stronger THz absorption.} 
\end{figure}

The system's imaging field of view is significantly enhanced by the implementation of a large-area atomic vapor cell. To demonstrate the advantage of this expanded field, two test objects were imaged: a two-inch brass disc engraved with the letters ``SARI'' (font width: 1.41 mm) and an aircraft model composed of zinc alloy and plastic measuring 35 mm $\times$ 28 mm. Fig. 4 shows both the physical photographs and the corresponding THz atomic fluorescence images of the objects, from which clear and distinguishable features can be observed. The results indicate that the system achieves an effective square imaging field of view of approximately 40 mm, which is currently limited by the aperture size of the imaging lens. It is anticipated that by employing a larger lens system with a 1:1 or higher magnification ratio, the full area of atomic fluorescence (approximately 50 mm $\times$ 50 mm) could be utilized for imaging. Furthermore, the overall imaging quality could be improved by optimizing the THz lens design and correcting optical aberrations in the imaging system. Although a thinner vapor cell (1 mm thickness) is used to reduce internal reflections, residual interference patterns caused by multiple THz reflections within the cell still degrade image quality. As the technology advances, this issue may be mitigated by fabricating even thinner vapor cells or by incorporating THz-compatible materials with high transmittance and anti-reflection coatings.

\subsection{Liquid Mixing Experiment and Analysis}
We employed the system to image the mixing process of two colorless liquids: deionized water and anhydrous ethanol. Anhydrous ethanol was first injected into a cuvette with an internal thickness of 1 mm, which was fixed at the object plane of the imaging system. Deionized water was then dripped naturally from above into the ethanol, and the entire process was recorded over a duration of 10 s using the sCMOS camera operating at 100 FPS. All captured images were processed using contrast enhancement algorithms and pseudo-color mapping. Representative frames from the video are shown in the upper panel of Fig. 5. From 0 to 120 ms, the water droplet freely falls above the ethanol surface. Between 120 ms and 250 ms, the droplet begins to merge with the ethanol, causing the liquid surface to rise. At approximately 250 ms, the droplet is fully integrated, and the surface reaches its highest position. Due to the significant difference in THz absorption between the two liquids, water absorbs THz radiation much more strongly than anhydrous ethanol. Therefore, the mixing process can be quantitatively analyzed by monitoring the average gray value beneath the liquid surface. In Fig. 5, we selected a 100 $\times$ 100 pixels square region just below the liquid surface (blue dashed box) and another region of the same size at the same vertical level but outside the THz field as a background reference (white dashed box). By computing the average gray values of both regions for each frame, we obtained their temporal evolution curves, as plotted in the lower panel of Fig. 5. After the droplet enters the ethanol (after 120 ms), the average gray value of the region beneath the surface steadily decreases, confirming the stronger THz absorption by water. These results further validate the system’s capability for high-speed THz imaging of dynamic processes.

We also conducted a preliminary investigation into the diffusion dynamics of water in ethanol. As illustrated in Fig. 6, each curve represents the average gray value at a different depth below the liquid surface (note that due to the inverted imaging setup, lower pixel values correspond to regions closer to the bottom of the ethanol solution). To reduce image noise, we applied a downsampling procedure by averaging every ten adjacent pixels in each column, thereby generating a single representative gray value per spatial segment. In the resulting curves, a downward trend is observed at several locations marked by black arrows, which can be attributed to the diffusion of water into these regions—causing a decrease in the local gray value due to the stronger THz absorption of water compared to ethanol. At a depth corresponding to 155 pixel, closer to the bottom of the cuvette, the gray value shows minimal change, indicating that the water had not yet diffused to that level within the observation window. 
Based on the data, we estimated the approximate diffusion velocity of water in ethanol. The cuvette has a width of 10 mm, which spans approximately 250 pixels in the horizontal direction of the image. This results in a spatial scale of 40 \textmu m per pixel. The observed diffusion front shifts from 295 pixel to 205 pixel, corresponding to a physical distance of approximately 3.6 mm. Given a time interval of 10 ms between the two frames and assuming uniform motion, the estimated diffusion speed is approximately 0.36 m/s under these experimental conditions. It should be noted that the actual diffusion behavior of water in ethanol is governed by complex intermolecular interactions and may involve non-uniform mixing dynamics. The above estimation serves only as a rough approximation to illustrate the system's capability. Nonetheless, these results demonstrate that the proposed THz imaging system can effectively visualize and quantify the mixing processes of colorless liquids with differing THz absorption characteristics. This highlights the potential of the technique to probe deeper physical parameters in liquid systems and offers a novel THz-based observational perspective for dynamic chemical or biological processes.

\section{Conclusion}
This study addresses the challenge of the limited imaging area in current THz imaging systems based on Rydberg atoms by proposing an optimization strategy involving a large-area atomic vapor cell and uniform laser beam shaping. By employing a 60 mm $\times$ 100 mm $\times$ 5 mm cesium vapor cell and incorporating Powell prisms, we effectively expanded the imaging field of view and improved the spatial uniformity of the atomic fluorescence distribution, without requiring significant increases in laser power or heating of the vapor cell. Experimental results demonstrate that the system is capable of real-time THz atomic fluorescence imaging over a large area (50 mm $\times$ 50 mm) with a spatial resolution of 1.25 mm, approaching the theoretical diffraction limit. The system was further applied to visualize and analyze the mixing process of two colorless liquids, demonstrating its potential for detecting polarity differences and observing dynamic diffusion processes in liquid systems. This work provides a valuable platform for high-resolution and high-sensitivity THz imaging, particularly in scenarios involving large-scale objects. It also broadens the scope of applications for Rydberg atom-based THz imaging.

Although certain limitations remain—including residual interference effects and component constraints, the continued development of high-transmittance THz materials, anti-reflection coatings, and high-power laser sources is expected to further enhance the performance and practicality of such systems. In the future, Rydberg atom-based THz imaging may become a powerful tool in biomedical diagnostics and advanced industrial inspection.

\section*{}


\bibliographystyle{IEEEtran}
\bibliography{citations}

\end{document}